\documentclass[runningheads]{llncs}
\usepackage[T1]{fontenc}
\usepackage{graphicx}
\usepackage{hyperref}
\usepackage{orcidlink}
\begin{document}
\title{Leveraging XP and CRISP-DM for Agile Data Science Projects}
%
%

\author{Andre Massahiro Shimaoka\inst{2,3}\orcidID{0000-0002-9400-8083} \and
Renato Cordeiro Ferreira\inst{1}\orcidID{0000-0001-7296-7091} 
 \and
Alfredo Goldman\inst{1}\orcidID{0000-0001-5746-4154}}

\authorrunning{Shimaoka et al.}
%
\institute{University of São Paulo, São Paulo, SP, Brazil \\
\and
Instituto de Pesquisas Tecnológicas, São Paulo, São Paulo, SP, Brazil \and
Federal University of São Paulo, São Paulo, SP, Brazil\\
\email{andre.shimaoka@unifesp.br}}

\maketitle              
\begin{abstract}
This study explores the integration of eXtreme Programming (XP) and the Cross-Industry Standard Process for Data Mining (CRISP-DM) in agile Data Science projects. We conducted a case study at the e-commerce company Elo7 to answer the research question: How can the agility of the XP method be integrated with CRISP-DM in Data Science projects? Data was collected through interviews and questionnaires with a Data Science team consisting of data scientists, ML engineers, and data product managers. The results show that 86\% of the team frequently or always applies CRISP-DM, while 71\% adopt XP practices in their projects. Furthermore, the study demonstrates that it is possible to combine CRISP-DM with XP in Data Science projects, providing a structured and collaborative approach. Finally, the study generated improvement recommendations for the company.


\end{abstract}
\section{Introduction}

Most research in Data Science focuses on technical resources, overlooking project organization and management. Many Data Science projects fail or fall short of delivering expected value, with 82\% of teams lacking a process model or methodology \cite{Saltz2019}. 

Cross Industry Standard Process for Data Mining (CRISP-DM) is a widely used process model in Data Science \cite{Martinez-Plumed2021}. It is technology-independent, adaptable across industries, and defines key steps for data science projects  \cite{Wirth2000}. However, it lacks predictability and does not follow agile principles and practices.\cite{Mariscal2010,Baijens2020}. 

Data science models are integrated into lines of code, and applying software engineering practices can enhance organization and efficiency in the development and maintenance of that code \cite{Cruz2017}. Agility is a concept that has been modestly explored in Data Science, presenting an opportunity to integrate the eXtreme Programming (XP) method with CRISP-DM \cite{Shimaoka2024}.

We conducted an empirical study in a real organizational context. To represent the field of Data Science, we used CRISP-DM as a reference for the process. To explore agility, we adopted the agile development method XP. The guiding question for this study is: 

\vspace{0.2cm}
\noindent
\fbox{
\begin{minipage}{11.7cm}
\vspace{0.3cm}
\textbf{RQ1: How can the agility of the XP method be integrated with CRISP-DM in Data Science projects?}
\vspace{0.3cm}
\end{minipage}
}

\subsection{CRISP-DM}
CRISP-DM is a structured process model used to guide data mining, data science, and analytics projects\cite{Martinez-Plumed2021}. It consists of six phases\cite{Chapman2000,Wirth2000}:
\vspace{-0.1cm}
\begin{itemize}
    \item \textbf{Business Understanding}: focuses on defining project objectives and business requirements, translating them into a data mining problem.
    \item \textbf{Data Understanding}: focuses on collecting and exploring the data to identify patterns, issues, and derive initial insights.
    \item \textbf{Data Preparation}: focuses on cleaning, selecting, and transforming the data to make it suitable for modeling.
    \item \textbf{Modeling}: focuses on selecting and applying appropriate modeling techniques, refining the model as necessary.
    \item \textbf{Evaluation}: focuses on assessing the model to ensure it meets business objectives and reviewing the steps taken.
    \item \textbf{Deployment}: focuses on implementing the model in the business environment, using systems, software, reports, or dashboards, with continuous monitoring to ensure alignment with business goals.
\end{itemize}

\subsection{eXtreme Programming}
Brief description of the XP practices used in this study:
\begin{itemize}
    \item \textbf{User Stories}: simple, short descriptions of features focused on user needs that guide development and help prioritize work \cite{BeckFowler2000,Cohn2004}.
    
    \item \textbf{Planning and Releases}: high-level plan covering work to be done each quarter, aligning goals and deliverables \cite{BeckFowler2000,Beck2004}.
    
    \item \textbf{Iterations}: short development cycles, usually 1 to 4 weeks, during which features are planned, built, tested, and small releases are made for continuous feedback\cite{BeckFowler2000,Beck2004}.
    
    \item \textbf{Slack}: buffer time to handle unexpected issues, allowing the team to maintain quality and avoid overload \cite{Beck2004}.
    
    \item \textbf{Whole Team}: a multidisciplinary team with all the skills necessary for the project \cite{Beck2004,Well2011}.
    
    \item \textbf{Informative Workspace}: a visual, organized workspace with clear, accessible information on project progress, promoting communication and transparency \cite{Beck2004,Well2011}.
    
    \item \textbf{Sustainable Pace}: a steady, healthy work rhythm without overloading, ensuring long-term productivity and quality \cite{Beck2004,Well2011}.
    
    \item \textbf{Pair Programming}: two developers work together at one station, reviewing and improving code simultaneously, enhancing quality and collaboration \cite{Beck2004}.
    
    \item \textbf{Continuous Integration}: frequent, automated code integration to ensure the software is always functional and minimize errors \cite{Fowler2006}.
    
    \item \textbf{Test-First Development}: writing automated tests before code to ensure requirements are met and code quality is high. This creates an efficient cycle of testing, coding, and refactoring \cite{Beck2004}.
    
    \item \textbf{Incremental Design}: system design done incrementally and iteratively, starting with simple solutions that are gradually refined, making future changes easier \cite{Beck2004,Fowler2007}.
    
    \item \textbf{Spikes}: quick exploration techniques to solve technical or design problems, evaluate potential solutions, and discard unfeasible ones \cite{Well2011}.
\end{itemize}

\section{Method}

\begin{figure}
\includegraphics[width=\textwidth]{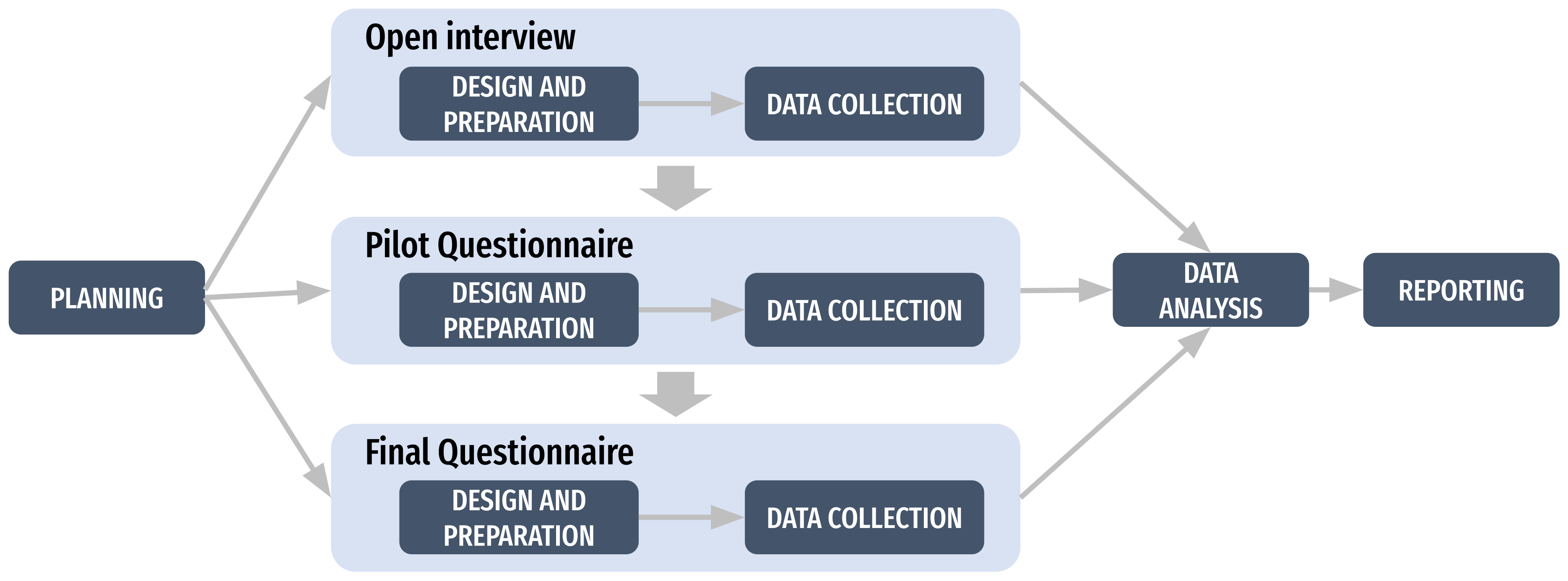}
\caption{Case study methodology} \label{figure1}
\end{figure}

In this study, We conducted a case study of Elo7, an e-commerce company founded in 2008 and a market leader in Brazil, specializing in handmade products, operating a platform with over 9 million items, and having an agile data science team. An intrinsic case study seeks to deeply understand an individual case in its specific aspects \cite{Ventura2007}. Furthermore, a revelatory case allows the researcher to examine a phenomenon previously underexplored in scientific research \cite{Yin2018}. Additionally, no relevant studies were found that address the combined use of CRISP-DM and XP \cite{Shimaoka2024}. Given that few companies adopt both methods, our objective was to understand a specific instance of the adoption of CRISP-DM and XP in Data Science at the company Elo7. \autoref{figure1} illustrates the case study process as applied in this study.

\subsection{Planning}
We defined the scope of the case study. We applied the convergence of evidence through interviews and questionnaires. This approach strengthens construct validity in case studies. Using multiple sources of evidence allows for diverse assessments of the phenomenon, improving the accuracy of the event presentation \cite{Yin2018}.
\subsection{Design and Preparation}

\subsubsection{Interview:} 
The open interview is used to explore a topic in depth, allowing the interviewee to freely discuss the proposed theme \cite{Bryman2016}.To understand the company and its processes in data science projects in an exploratory manner, we decided to conduct a qualitative, open-ended, and unstructured, which provided input for the Survey.

We prepared the open interview to be conducted remotely (via telepresence) with the data leadership of the company, focusing on the specific topic of Agility in Data Science. We divided the interview into a discussion on the general topic of agility in data science, followed by an exploration of the aspects of CRISP-DM and XP.

\vspace{-0.3cm}
\subsubsection{Questionnaire:}
To guide the evidence collection through questionnaires, three main aspects were analyzed: (1) the application of the CRISP-DM stages by the data science team, (2) the adoption of XP practices within the team, and (3) the specific XP practices used at each stage of the CRISP-DM framework in the company.

Next, we structured the survey with closed questions that utilize a 5-point Likert scale, which enables precise and standardized data collection that respondents can easily understand. The response options are: \textit{Never}, \textit{Rarely}, \textit{Occasionally}, \textit{Frequently}, and \textit{Always or More frequently}. To collect data on the application of the CRISP-DM stages by the data science team, we designed a questionnaire with 20 questions divided across the six phases of CRISP-DM: \textit{Business Understanding}, \textit{Data Understanding}, \textit{Data Preparation}, \textit{Modeling}, \textit{Evaluation}, and \textit{Deployment}. These questions cover data science activities, ranging from business vision and strategic alignment to model deployment, including technical tasks such as data preparation and modeling.

To assess the implementation of XP practices within the team, a questionnaire was designed with 13 questions covering the following XP practices: \textbf{User Stories}, \textbf{Releases}, \textbf{Iterations}, \textbf{Slack}, \textbf{Whole Team, Informative Workspace}, \textbf{Sustainable Pace}, \textbf{Pair Programming}, \textbf{Continuous Integration}, \textbf{Test-Driven Development}, \textbf{Incremental Design}, and \textbf{Spikes}. Both the CRISP-DM and XP questions included an optional open-ended field to collect additional information. Finally, to assess XP practices across CRISP-DM stages, participants linked each XP practice to one or more CRISP-DM phases. The entire questionnaire can be found in the supplementary material (https://.....).

\subsection{Data Collection}

\subsubsection{Interview:} 
On November 22, 2022, we held a meeting with the lead data scientist at Elo7 from 5:00 p.m. to 6:45 p.m. via Google Meet. We presented the CRISP-DM process model, and the data scientist confirmed that all stages and tasks in the model aligned with the practices at Elo7. She also reported that the company applied the agile XP method and executed Data Science activities in iterative cycles.

She emphasized the use of the Spikes practice, which the team used to solve technical challenges and gain a deeper understanding of the problem and business objectives. Additionally, the company routinely conducted both product and process reviews. Finally, Elo7 deployed models through software or applications and maintained a model training platform, which facilitated the maintenance and improvement of deployed models.

\subsubsection{Questionnaire:}
Between January 16 and 20, 2023, we conducted a pilot with three Data Science professionals from Elo7 to validate the format and clarity of the questions. Based on the feedback, we made the following adjustments: the inclusion of a figure depicting the phases of CRISP-DM, an improvement in the description of XP practices, and the replacement of the Likert scale of agreement with a frequency scale.

Subsequently, we distributed the questionnaire to all professionals involved in Data Science projects at Elo7. It was available from February 6 to 23, 2023, on the Google Forms platform, administered online, unsupervised, with instructions for completion. The form provided information on data confidentiality and anonymity, and all respondents agreed to the terms. No personal data was collected.

\section{Results}

The survey included 13 professionals, encompassing roles such as data scientists, machine learning engineers, and product managers. This diversity of profiles enabled a comprehensive analysis of the practices and challenges encountered in the application of CRISP-DM and XP.

\subsection{CRISP-DM}

\begin{figure}
\includegraphics[width=\textwidth]{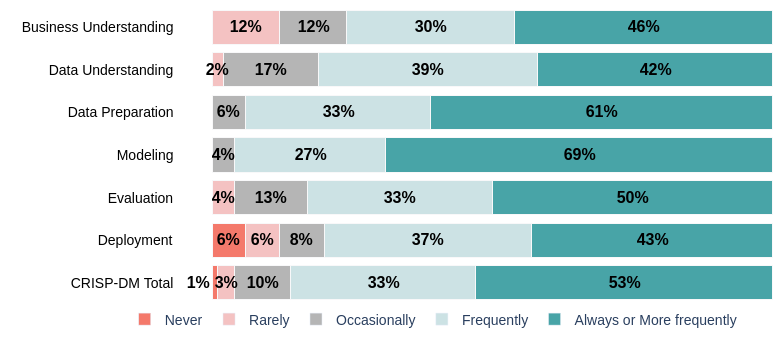}
\caption{Survey results on the application of CRISP-DM stages by the agile data science team} \label{figure2}
\end{figure}

The results presented in \autoref{figure2} indicated a high level of adherence to the use of CRISP-DM, with 86\% of responses indicating usage between frequently and always. This demonstrated that CRISP-DM was widely adopted by the majority of respondents, with consistent application in their activities. Only 10\% applied CRISP-DM occasionally, while 4\% rarely or never used it.

\subsubsection{Business Understanding:}
Part of the team reported a lack of alignment with business objectives. Additionally, the focus was generally on data exploration rather than on business goals. This phase was often led by product managers, with limited involvement from data scientists.

\subsubsection{Data Understanding:} 
The company had well-established and previously explored databases. Additionally, hypotheses from prior projects reduced the need for these activities. When new hypotheses were required, both technical team members and product managers should participated, ensuring alignment with business objectives and goals.

\subsubsection{Data Preparation:} 
Activities such as removing inconsistent records, formatting data, and aggregating similar attributes may occur concurrently with the Modeling phase. The CRISP-DM framework allows for overlap between the Data Preparation and Modeling phases\cite{Chapman2000}, which can create the impression that Data Preparation activities are being carried out within the Modeling phase. Some of the team confused these activities and regarded them as part of the Modeling phase instead of Data Preparation.

\subsubsection{Modeling:} 
This was the most well-known and widely adopted phase among data science team.

\subsubsection{Evaluation:} 
The evaluation process was well known and practiced by the data science team. However, part of the model evaluation included A/B testing, which was conducted during the Deployment phase. This experimental approach presents two versions of the same element to different groups of users randomly, aiming to determine which version performs better against a business metric \cite{Kohavi2017}.

\subsubsection{Deployment:} 
In this phase, ML engineers were actively involved in contingency planning and monitoring, while data scientists contribute less, which explained data scientists never or rarely performing these activities. Data scientists mainly focused on data preparation and modeling, while ML engineers concentrated on deploying infrastructure and systems that support the model, including model monitoring.

\subsection{XP practices}

\begin{figure}
\includegraphics[width=\textwidth]{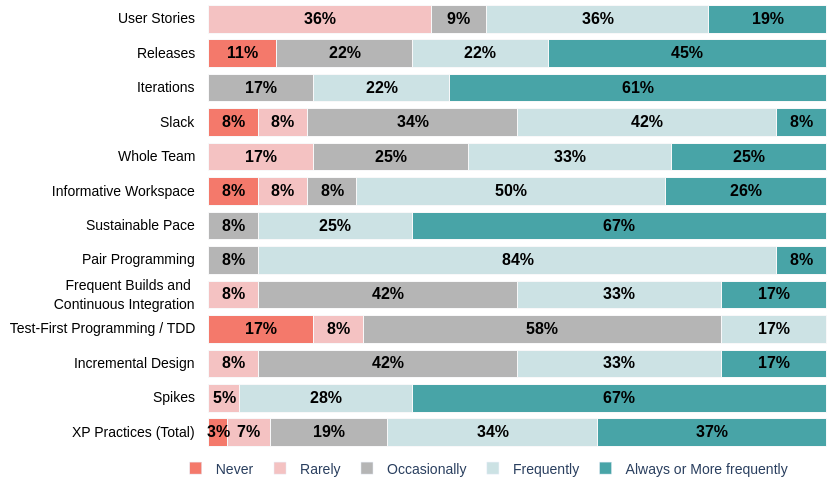}
\caption{Survey results on XP Practices adoption by the agile data science team} \label{figure3}
\end{figure}

The results presented in \autoref{figure3} indicated that 71\% of responses reflected the use of agile XP practices between frequently and always. Additionally, 19\% applied them occasionally, while 10\% rarely or never use them. This suggested that agile XP practices were widely adopted, although a portion of respondents either used them sporadically or not at all, which may point to areas for improvement in their integration or awareness within the organization.

\vspace{-1cm}
\subsubsection{User stories:} 
In the company, user stories were not always clear or understandable to product managers. The focus was often on technical terms, sometimes overlooking business rules. There was also confusion about the responsibilities for writing user stories and tasks, with many stories being written at a technical task level.

\subsubsection{Releases:} 
Some ML engineers and data scientists were not familiar with how this process was applied internally within the company. Although the company had a long-term product roadmap, it was not updated quarterly as in XP release planning \cite{Beck2004}. The roadmap maps the vision and direction of releases and outlines how the product portfolio will meet business objectives \cite{Munch2019}. A gap existed between the technical team and the strategic side of the company regarding this practice and the long-term vision. 

\subsubsection{Iterations:} 
In the company, iterations typically lasted for two weeks. There was a difference in how delivery was perceived within iterations among data science professionals. Managers viewed deliveries as ready-to-use outputs for users, while data scientists and ML engineers focused more on finished features, though not always available for use. Exploratory analyses did not result in working software, but they still added value to users through reports, insights, graphs, and other deliverables.

\subsubsection{Slack:} 
Smaller tasks, which can be postponed, are included to handle unexpected issues \cite{Beck2004}. However, the tasks chosen for the iteration were essential for delivery, leaving no buffer, and slack practices were not always used. Additionally, the company encouraged extra project activities, such as study time and hours for personal use, but not everyone recognized them as part of slack practices.

\subsubsection{Whole Team:} 
Certain skills and roles, like front-end developers and data engineers, were missing from the team. As a result, they depended on external members for specific tasks.

\subsubsection{Informative Workspace:} 
Most team members used dashboards and charts to display relevant project information. This approach provided stakeholders with views of project progress and highlighted any obstacles encountered.

\subsubsection{Sustainable Pace:} 
The company fostered a sustainable work culture by minimizing unproductive overtime and preventing employee overload.

\subsubsection{Pair programming:} 
For the data science team, it was an effective practice for sharing knowledge and turning ideas into code. This applied to both architecture design and coding, as well as exchanging ideas in exploratory data science activities. However, when done for long periods and frequently, it can become exhausting \cite{Williams2002}.

\subsubsection{Frequent Builds and Continuous Integration:} 
For ML engineers, this practice required more robust infrastructure. Additionally, the lack of maturity in automating continuous integration processes made it difficult to fully adopt these practices. For data scientists, automated tests were rare during experimentation and data analysis, but more common during model deployment.

\subsubsection{Test-First Programming:} 
This practice was the least adopted XP practice among the data science team. It is challenging to implement in analytical activities, while it is more commonly used during the development of software that supports the models. 

However, this practice can be applied during algorithm implementation, including tasks such as data preparation, model creation, tuning, training, and testing. This extends beyond the deployment phase. Therefore, it is necessary to train data scientists in this practice to ensure its effective adoption.

\subsubsection{Incremental Design:} 
For managers, solutions in the first iteration often had high complexity, making incremental design challenging. It is important to avoid overly complex solutions or those that require more resources than necessary. When building models, selecting a small number of parameters that are easier to interpret and explain is essential \cite{Lazar2010}.

\subsubsection{Spikes:} 
This practice was the most commonly used practice by the team. Spikes were primarily used to find answers to challenging problems and explore potential solutions.

\subsection{Combining the XP practices with CRISP-DM phases}

\begin{figure}
\includegraphics[width=0.9\textwidth]{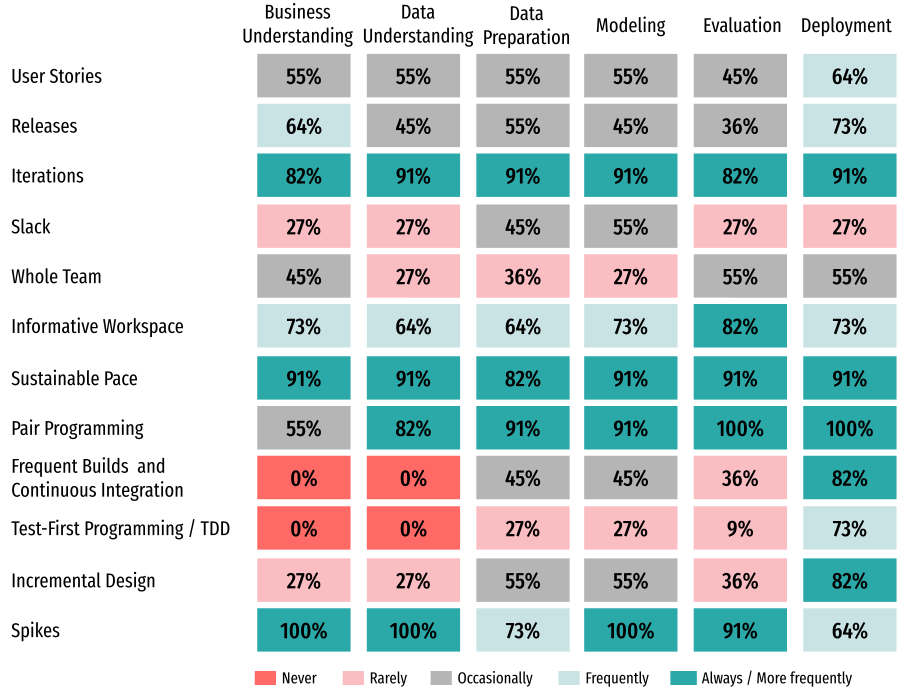}
\caption{Survey results on the integration of CRISP-DM and XP Practices by the Agile Data Science Team} \label{figure4}
\end{figure}

An XP practice can relate to multiple phases, just as a CRISP-DM phase can involve several XP practices. \autoref{figure4} showcases the relationship between CRISP-DM phases and XP practices. 

\subsubsection{User Stories with CRISP-DM Phases:} 

The team recognized that user stories could be created, refined, or consulted at any phase of CRISP-DM, though they used them less frequently in the \textit{Evaluation} phase. In this phase, models are evaluated from the perspective of business goals \cite{Chapman2000}, and user stories help ensure alignment with these objectives \cite{Cohn2004}. However, the technical language in user stories impacted their use in this stage by the team.

\subsubsection{Releases with CRISP-DM Phases:} 
Releases practices were more common in the \textit{Business Understanding} and \textit{Deployment phases}, as these phases involved medium- and long-term planning and product delivery according to the plan. In the more analytical phases, such as \textit{Data Understanding}, \textit{Data Preparation}, and \textit{Modeling}, their use was less frequent. However, the team understood that the Release plan can be created, consulted, refined, and updated at any stage of CRISP-DM.

\subsubsection{Iterations with CRISP-DM Phases:} 
Iteration practices, including planning, using short cycles, and delivering in small increments, frequently integrated into all phases of CRISP-DM.

\subsubsection{Slacks with CRISP-DM Phases:} 
Slacks practices were seldom adopted in the phases of CRISP-DM. Extraproject activities and free hours, such as time for training and skill development, were not linked to any specific phase. Additionally, data science activities allowed little buffer for unforeseen events

\subsubsection{Whole Team with CRISP-DM Phases:} 
During the \textit{Business Understanding} phase, the technical team (data scientists and ML engineers) had limited participation, while in the other phases, the business team (product managers) was minimally involved. As a result, the team recognized the need for a more integrated approach, with active participation from everyone in all phases of CRISP-DM.

\subsubsection{Informative Workspace with CRISP-DM Phases:} 
The team frequently used dashboards and charts to display progress, risks, and project issues throughout all stages of CRISP-DM.

\subsubsection{Sustainable Pace with CRISP-DM Phases:} 
The sustainable pace practice is frequently adopted in all phases of CRISP-DM, with minimal overtime, focusing on productivity while also caring for the teams well-being.

\subsubsection{Pair Programming Pace with CRISP-DM Phases:} 
The pair programming practice was widely adopted across the phases of CRISP-DM. The team used it not only in technical activities but also in other activities such as requirements definition

\subsubsection{Frequent Builds and Continuous Integration with CRISP-DM Phases:} 
This practice was rarely used by the team, even when it involved model creation and data processing, which required code manipulation. The team adopted this practice frequently only in the software development phase that utilizes the models, specifically during the \textit{Deployment} phase.

\subsubsection{Test-First Programming with CRISP-DM Phases:} 
The team used the Test-First Programming practice the least. However, they frequently applied it in the \textit{Deployment} phase. Data scientists showed resistance to this approach when creating models, preparing data, and analyzing data, even when code manipulation was involved.

\subsubsection{Incremental Design with CRISP-DM Phases:} 
This practice was more frequent in the \textit{Deployment} phase. In the \textit{Modeling} phase, the models were created with high complexity. Additionally, in the \textit{Business and Data Understanding} phases, little importance was given to solution design. However, the culture of starting with a simple solution and gradually incrementing it should have been adopted from the very beginning. \cite{Fowler2007}.

\subsubsection{Spikes with CRISP-DM Phases:} 
Spikes practices were used in all phases of CRISP-DM. In the \textit{Business Understanding} and \textit{Data Understanding} phases, the requirements, data, and problem definition were still immature. Therefore, the practice helped to explore insights and problems. In the Modeling phase, it was consistently applied to explore, compare, and discard various algorithms and models.

\subsection{Recommendations}
The adoption of XP and CRISP-DM in data science projects can be challenging. Issues such as misinterpreted user stories, lack of transparency in the product roadmap, discrepancies in team deliverables, lack of recognition of slack practices, non-multidisciplinary teams, infrequent testing, low maturity in continuous integration, and high model complexity highlight the need for targeted recommendations.

The list below presents potential recommendations for the company:
\begin{itemize}
    \item Stories should be clear and made available to the entire team;
    \item Stories should be written in business language and perspective;
    \item Provide training for the data science team on writing user stories;
    \item Hold workshops to present the strategic planning and roadmap to the entire team;
    \item Align the definition of deliverables in data science;
    \item Align the concept of slack time within the team Extra-project activities (e.g., studies and free time) can be considered slack time practices;
    \item Assess the possibility of integrating new roles or training the team, such as front-end developers and data engineers;
    \item Invest in MLOps \cite{Testi2022} to improve the maturity and frequency of XP practices, such as frequent builds and continuous integration;
    \item Train and encourage data scientists to use software engineering practices, such as testing, frequent builds and continuous integration;
    \item Train the team in Behavior-Driven Development (BDD) \cite{Smart2014} sing a common language understandable by data scientists, product managers, and ML engineers, promoting testing practices.
    \item Encourage the data science team to think of solutions that start simpler (less complex).
\end{itemize}

\subsection{Threats to Validity}
The study was conducted within a single company, which limited the ability to generalize the results to other organizations or sectors. Additionally, the company had an organizational culture that favored the adoption of agile methodologies, which may not apply to companies facing greater resistance to change.

Another limitation was that the study relied primarily on interviews and surveys regarding the perceptions of the team. Due to confidentiality concerns, no documentary evidence was analyzed to verify the adoption of the methods.

\section{Conclusion}

This study aimed to explore how the agility of the XP method can be integrated with CRISP-DM in data science projects, addressing the research question: \textit{\textbf{How can the agility of the XP method be integrated with CRISP-DM in data science projects?}} To achieve this, three aspects were analyzed: (1) the application of CRISP-DM stages by the data science team, (2) the adoption of XP practices within the team, and (3) the specific XP practices applied at each stage of the CRISP-DM framework.

The data science team at Elo7 applied all CRISP-DM stages frequently, with the \textit{Modeling} and \textit{Data Preparation} phases being the most adopted. However, the team faced challenges differentiating activities from different phases that occurred simultaneously. Additionally, data scientists needed to be involved in more strategic phases, such as \textit{business understanding}.

The team adopted all XP practices, with practices like \textbf{Test-First Programming}, \textbf{Slack}, and \textbf{User Stories} being applied less frequently. On the other hand, \textbf{Sustainable Pace}, \textbf{Spikes}, \textbf{Iterations}, and \textbf{Pair Programming} were the most frequently applied practices. However, challenges and difficulties in adopting XP were identified, leading to recommendations for the team.

In conclusion, the study showed that the core XP practices could be applied in any CRISP-DM phase, except for \textbf{Frequent Builds}, \textbf{Continuous Integration}, and \textbf{TDD}, which were not applicable in the \textit{Business Understanding} and \textit{Data Understanding} phases. However, some practices, such as \textbf{Slack} and \textbf{Whole Team}, were rarely applied.

This work empirically reported the use of the CRISP-DM process model and XP practices within an organizational environment. Based on the results, it can be concluded that combining CRISP-DM and XP can be a viable approach for agile data science projects. Furthermore, the results suggest that agile practices can be implemented without compromising the data science process.

The case study provided insights into the use of the CRISP-DM process model and XP practices in an e-commerce company. Further studies in different companies and among data science professionals with varied experience levels would help assess agility in data science and identify challenges and benefits of using these practices.

Finally, it would be valuable to explore how CRISP-DM and XP can be adapted for larger and distributed teams (more than 13 people). Such a study could assess how these practices can be scaled to meet exploratory and specific needs in data science projects.

\begin{credits}
\subsubsection{\ackname} 

\subsubsection{\discintname}
The authors have no competing interests to declare that are relevant to the content of this article.
\end{credits}
%
%
%
%

\end{document}